\documentclass[prb,twocolumn,showpacs]{revtex4}
\usepackage{graphicx}
\usepackage{times}
\usepackage{bm}
\usepackage{amsmath, verbatim}

\def\sgn{{\text{sgn\,}}}
\def\be{\begin{equation}}
\def\ee{\end{equation}}
\def\bea{\begin{eqnarray}}
\def\eea{\end{eqnarray}}
\def\bse{\begin{subequations}}
\def\ese{\end{subequations}}

\def\sgn{{\text{sgn\,}}}

\def\be{\begin{eqnarray}}
\def\ee{\end{eqnarray}}

\newcommand{\ua}{\uparrow}
\newcommand{\da}{\downarrow}

\newcommand{\expect}[1]{\langle {#1} \rangle}

\begin{document}

\title{Probing non-Abelian statistics with Majorana fermion interferometry in spin-orbit-coupled semiconductors}
\author{Jay D. Sau$^1$}
\author{Sumanta Tewari$^{2}$}
\author{S. Das Sarma$^1$}
\affiliation{$^1$Condensed Matter Theory Center and Joint Quantum Institute, Department of Physics, University of
Maryland, College Park, Maryland 20742-4111, USA\\
$^2$Department of Physics and Astronomy, Clemson University, Clemson, SC
29634}

\begin{abstract}

The list of quantum mechanical systems with non-Abelian statistics has recently been
expanded by including  generic spin-orbit-coupled semiconductors (\emph{e.g.}, InAs) in proximity to a $s$-wave superconductor.
Demonstration of the anyonic statistics using Majorana fermion interferometry in this system
is a necessary first step towards topological quantum computation (TQC).
However, since all isolated chiral edges that can be created in the semiconductor are charge neutral, it is not clear if electrically controlled
interferometry is possible in this system. Here we show that when two isolated chiral Majorana edges are brought into close contact, the resultant
 interface supports charge current, enabling electrically controlled Majorana fermion interferometry in the semiconductor structure.
 Such interferometry experiments on the semiconductor are analogous to similar interferometry experiments
 on the $\nu=5/2$ fractional quantum Hall systems and on the surface of a 3D strong topological insulator, illustrating the
 usefulness of the 2D semiconductor heterostructure as a suitable TQC platform.
In particular, we proposed Majorana interferometers may be the most direct 
method for establishing non-Abelian braiding statistics in 
topological superconductors.

\end{abstract}

\maketitle


\section{Introduction}

Topological many body systems, characterized by exchanges of the
 particles unitarily rotating the wave function in the degenerate 
ground state subspace (non-Abelian statistics),
are called non-Abelian quantum systems.~\cite{kitaev1,nayak_RevModPhys'08}
Such systems
have been proposed as a fault-tolerant platform for topological quantum computation (TQC).~\cite{nayak_RevModPhys'08,dassarma_prl'05}
Some key, albeit exotic, condensed matter systems, such as the Pfaffian states in fractional quantum Hall (FQH) systems~\cite{dassarma_prl'05,Read_prb'00} and chiral $p$-wave superconductors/superfluids,~\cite{ivanov_prl'01,DasSarma_PRB'06,tewari_prl'2007} as well as the surface state of a 3D strong topological insulator (TI),~\cite{fu_prl'08} have been identified as non-Abelian systems.
 Recently, this list has been expanded by showing that even a regular semiconducting thin film (\emph{e.g.}, InAs), proximate to an ordinary 
 $s$-wave superconductor and a magnetic insulator, can support non-Abelian excitations. \cite{sau_prl'09,annals,long}
 Since neither special materials nor exotic physics are needed to produce the non-Abelian excitations, this structure is potentially one of the simplest to probe non-Abelian quantum matter and therefore has attracted
considerable attention in the recent literature. \cite{alicea,franz,roman,oreg,qi,linder,blueprint,njp,hassler,universal,roman-tudor1,roman-tudor2,chiral_domain,hole-doped,critical}

 The non-Abelian excitations in the semiconductor in this proposal, 
 are the zero-energy Majorana fermions, defined by the self-hermitian operators $\gamma_0; \gamma_0^{\dagger}=\gamma_0$,
 localized at vortices or the sample edges induced by the superconducting
proximity effect. \cite{sau_prl'09}
 Even though the non-Abelian excitations in the $\nu=5/2$ FQH system are 
 charged, the Majorana excitations in the semiconductor are charge-neutral.
 The charge neutrality of the Majorana excitations makes them
 unresponsive to an applied electric field and therefore 
difficult to detect experimentally. On the other hand,
 electrically controlled edge state interferometry of the Majorana
 excitations \cite{dassarma_prl'05,stern_prl'06,bonderson_prl'06}
 is a key experiment both to probe their exotic statistics as well
 as to implement TQC. \cite{bravyi1,bravyi2,bonderson_ann_phys}
 Recently it has been shown that on the surface of 3D TI (in which
 the Majorana excitations are also charge-neutral)
 an interface between two domains with opposite directions of
 magnetization (but no superconductivity)
 supports gapless \emph{charged} Dirac fermionic modes.
 \cite{fu_prl'09,akhmerov_prl'09}
 These charged modes can break up into two neutral Majorana edge modes
 in the presence of superconductivity.
 Since the charged fermion current can be
 controlled by external electric potentials, this
 provides a way to electrically control the neutral current of the edge state Majorana fermions in a TI, and thus to detect them.

  For a semiconductor, however, the above method does not work because
 there is no well-defined edge mode
  at the boundary between two domains with opposite directions of
 magnetization.
  This is because, unlike the TI surface, the system remains
 gapless even with non-zero magnetization
 in the absence of superconductivity.\cite{sau_prl'09}
 In the presence of superconductivity, however,
  all isolated chiral edge modes (\emph{e.g}., edges 1 and 2 in Fig. 1) that can be created
  in this system are Majorana modes, hence charge-neutral.
Very recently a set of proposals \cite{grosfeld1,grosfeld2} atempt 
to avoid this problem by
 using the transport of Majorana fermions in vortex cores of
 superconductors
to create measurable interferometry of vortices in superconductors.
The original proposals \cite{grosfeld1} for such interferometry suffer
 from the possibly
large vortex mass in superconducting systems making the observation of
quantum interference in such
proposals unlikely. The more recent proposals \cite{grosfeld2}
using Josephson vortices instead of Abrikosov vortices
can possibly  overcome the problem of a large vortex mass but might
still face the problem of a very small minigap. Therefore, electrically controlled edge state interferometry
of Majorana excitations remains a crucially important experiment to probe the non-Abelian
statistics of Majorana fermions and implement TQC
on $2D$ spin-orbit coupled semiconductor systems.
This is the problem we tackle in this work.

In this paper, we start with the existence of chiral Majorana
 edge modes and Majorana fermions shown in Refs.[\onlinecite{sau_prl'09,annals,long}]
and propose an interferometry experiment to demonstrate the
non-Abelian character of these charge neutral Majorana fermions.
To this end we first show that chiral Majorana modes located at the
 isolated edges in the
semiconductor (edges 1 and 2 in Fig. 1 with the separation $W$ large)
 are indeed charge-neutral. However, when the separation $W$ is on the order of the coherence length $\xi$ or smaller,
 the wave functions localized at the edges overlap and the resulting mode acquires charge.
 In the limit of zero separation between the edges, it is an interface separating regions with
 opposite directions of magnetization. We show by explicit calculations that even though such an interface
can be thought of as created by
 superposing two originally charge-neutral Majorana edge modes,
 it carries a chiral quasiparticle charge current which can be controlled by external bias voltages. The reason
such a charge current can arise
although the Majorana modes themselves are neutral is, of course, the
presence of superconductivity in some sense ``violates'' the naive
charge conservation because of the spontaneous breaking of the U(1)
symmetry and the existence of phase coherent Cooper pairs.
Recently, similar charged quasiparticle modes have been argued to exist at the domain
walls of chiral $p$-wave superconductors. \cite{chiral_domain}
 We give two specific interferometer designs suitable
 for testing non-Abelian statistics of quasiparticles and implementation
 of TQC.

In Sec. II we describe the basic BdG Hamiltonian for treating the topological superconducting properties of the generic semiconductor-superconductor 
sandwich structure proposed in Sau et. al. \cite{sau_prl'09}, 
discussing the condition for the emergence of the non-Abelian Majorana 
mode. In Sec. III we discuss the magnetic domain wall where the chiral 
Majorana modes fuse to produce chiral Dirac fermion modes. In Sec. IV we 
analytically study the charge conductance of the chiral Dirac modes 
in the wide domain wall limit in which case it is possible to 
think of the domain wall as composed of a pair of weakly overlapping 
chiral Majorana modes. We check numerically in Sec. V, that the
 qualitative features of the wide domain wall limit survive in the 
narrow domain wall limit, finally describing our proposed 
interferometer in Sec. VI. We conclude in Sec. VII summarizing our 
results and discussing some open questions.  

\section{BdG Hamiltonian for sandwich structures}
  We consider the BdG Hamiltonian for a semiconductor (Sm) in which an
 $s$-wave superconducting pair potential $\Delta$ and
  a Zeeman splitting $V_Z$ are induced
 by proximity effect from a superconducting layer (SC) and a ferromagnetic
insulator layer (F)(Fig. 1). This has been shown to be possible
experimentally \cite{merkt,exchangebias} and theoretically.~\cite{tudor,long}
Below, we review the excitation spectrum and derive an effective Hamiltonian
 of a semiconductor thin film sandwiched between a layer of SC and F.
The SC - Sm - F heterostructure, is described by a model defined by the Hamiltonian
\begin{equation}
H_{\rm tot} = H_{\rm{Sm}} + H_{\rm{SC}} + H_{\rm{F}} + H_{\tilde{t}_{\rm{SC}}} + H_{\tilde{t}_{\rm{F}}} \label{Htot}
\end{equation}
where $H_{\rm{Sm}}$, $H_{\rm{SC}}$ and $H_{\rm{F}}$ are the Hamiltonians describing the
Sm, SC and F layers respectively. $H_{\tilde{t}_{\rm{SC}}}$ and
$H_{\tilde{t}_{\rm F}}$ represent the tunneling Hamiltonians at the Sm-SC and Sm-F interface respectively.

The low-energy Hamiltonian for the Rashba SO-coupled Sm layer, $H_0$,
is given by
\begin{equation}
H_{\rm{Sm}}({\mathbf k}) =\sum_{\bm k}\psi_{\rm{Sm}}^\dagger[\frac{k^2}{2m^*} -\mu+\alpha (\bm\sigma\times \bm k)\cdot\hat{\bm z}] \psi_{\rm{Sm}} \label{H0eff}
\end{equation}
where $m^*$ is an effective mass.
The ferromagnetic insulator $F$ is described by the Hamiltonian
\begin{equation}
H_{\rm{F}}=\sum_{\bm k}\psi^\dagger_{\rm{F}}(\epsilon_k-\mu_{\rm{F}}+V_{\rm{F}}\sigma_z)\psi_{\rm{F}}
\end{equation}
 with a mean-field Zeeman order parameter $V_{\rm{F}}$ and a
 chemical potential $\mu_{\rm{F}}$ in the  Zeeman spin-split gap of the
ferromagnetic insulator. Similarly, the $s$-wave superconductor
can be described by a mean-field BCS Hamiltonian
\begin{equation}
H_{\rm {SC}}=\sum_{\bm k}\psi_{SC}^\dagger(\epsilon_k-E_F)\psi_{SC}+\Delta_{SC}\psi^\dagger_{SC}\psi^\dagger_{SC}.
\end{equation}
  Here $\psi_{\rm{SC}}^\dagger$, $\psi_{\rm{Sm}}^\dagger$ and $\psi_{\rm{F}}^\dagger$ are the relevant electron spinor creation operators.
In the absence of tunneling between the layers, Sm is
 normal with no superconductivity.
The tunneling term
\begin{equation}
H_{\tilde{t}_{SC}}=\tilde{t}_{\rm{SC}}\psi^\dagger_{\rm{SC}}\psi_{\rm{Sm}}+h.c
\end{equation}
 which transfers electrons between the Sm-SC layers leads to a finite
value for the  superconducting order parameter
 $\langle \psi_\sigma(r)\psi_{\sigma'}(r')\rangle$
 on the Sm surface by the proximity effect. \cite{degennes}
 Similarly,  the tunneling term
\begin{equation}
H_{\tilde{t}_{F}}=\tilde{t}_{\rm{F}}\psi^\dagger_{\rm{F}}\psi_{\rm{Sm}}+h.c
\end{equation}
leads to a proximity-induced ferromagnetic order parameter
 in the Sm. (In some situtations, the magnetic insulator can be 
replaced by an external magnetic field -- the main necessary ingredient is
 an effective spin splitting in the Sm layer without any magnetic 
field induced magneto-orbital effects. )

\begin{figure}
\centering
\includegraphics[scale=0.3,angle=0]{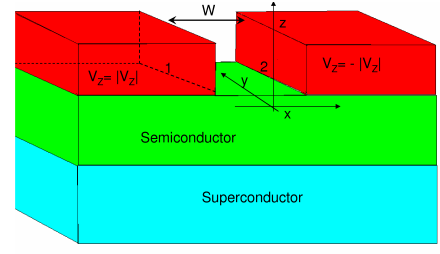}
\caption{Structure of an interface along y. Magnetic insulators shown as (red) top layer induce opposite
Zeeman splitting in the semiconductor. Edges $1$ and $2$, each separating regions of non-Abelian and
Abelian $s$-wave superconductors, are separated by interface width $W$. Such edges, when isolated
($W \gg \xi$, with $\xi$ the coherence length), carry charge-neutral chiral Majorana modes.}\label{Fig1}
\end{figure}

 We start by considering the ferromagnetic proximity effect induced in the
Sm layer from the F-Sm interface when the states of the lowest
 semiconductor band with wave vectors near ${\mathbf k}=0$
 have energies inside the insulating gap of the ferromagnetic insulator F.
 The insulating degrees of freedom of the ferromagnetic insulator, $F$, can be integrated out and replaced by an
 interface self-energy. When projected onto the low-energy subspace,
 this contribution becomes
\begin{equation}
\Sigma_{\sigma\sigma^{\prime}}^{(F)}({\mathbf k}, \omega) = -|\tilde{t}_F|^2 \vert\psi_{\mathbf k}(z_0)\vert^2 G_{\sigma\sigma^{\prime}}^{(F)}({\mathbf k}, \omega; z_0), \label{Sigmam}
\end{equation}
where $\tilde{t}_F$ is proportional to the transparency of the F-Sm
interface, $G_{\sigma\sigma^{\prime}}^{(F)}({\mathbf k}, \omega; z_0)$
 is the Green function of the ferromagnetic insulator (F)
 and $\vert\psi_{\mathbf k}(z_0)\vert^2$ the amplitude of the semiconductor wave function, both at the interface.
Note that the imaginary part of the Green function vanishes for values of $\omega$ within the insulating gap. Also, because the energies of interest are much smaller than the insulator bandwidth, $\omega\ll\Lambda_{\sigma}$, we can neglect the frequency dependence in Eq. (\ref{Sigmam}).
 From Eq. (\ref{Sigmam}) one immediately notices that, neglecting such
dynamical effects, i.e., setting $\omega=0$,
the proximity effect self-energy, is equivalent to
 an effective Zeeman splitting
\begin{equation}
V_Z=\frac{\Sigma_{\ua\ua}({\mathbf k}\sim 0, \omega=0)-\Sigma_{\da\da}({\mathbf k}\sim 0, \omega=0)}{2}\label{Vz}
\end{equation}
which creates a gap in the nth semiconductor band near ${\mathbf k}=0$
that is proportional to the amplitude of the wave function at the
 interface times the square of the interface transparency.

 Thus the dispersion of the low-energy bands in the Sm
 layer can be obtained by adding the
 Zeeman splitting (Eq.~\ref{Vz}) to the effective theory described
 by Eq. (\ref{H0eff}) to obtain a low-frequency and long-wavelength (i.e $\omega\sim 0$ and $\bm k\sim 0$) effective Hamiltonian
\begin{equation}
H_{Sm}({\mathbf k}) =\frac{k^2}{2 m^*} -\mu+\alpha (\bm\sigma\times \bm k)\cdot\hat{\bm z} +V_Z\sigma_z, \label{Heff}
\end{equation}
with an effective Zeeman splitting $V_Z$ which is determined by the
microscopic tunneling parameters at the F-Sm interface and
 can be tuned by: a) modifying the semiconductor film thickness,  b) applying a bias voltage, and c) changing the semiconductor - MI coupling.\cite{long}. Explicit calculations
 for various sets of parameters show that this low-energy theory
 represents an excellent approximation for all ${\mathbf k}$ values
of interest.\cite{long}

Next, we turn our attention to the effects induced by the proximity
 of an s-wave superconductor on the Sm-F heterostructure.
 We consider a semiconductor thin film and create a new interface at
 the free surface of the semiconductor by coupling it to a SC with an
 s-wave gap $\Delta$. Below we argue that, turning on coupling to
 the SC opens up a small gap near the fermi surface of the electron-doped
semiconductor described by the Hamiltonian in Eq.~\ref{Heff}.\cite{long}

In order to understand this behavior, it is useful to develop an
 effective low-energy theory for the SC proximity effect that is
 analogous to the
one discussed for the magnetic proximity effect. Thus the effect of the
superconductor can be described by an
effective self-energy \cite{robustness,tudor}
\begin{equation}
\Sigma_{\sigma\tau;\sigma^{\prime}\tau^{\prime}}^{(SC)}({\mathbf k}, \omega) = -|\tilde{t}_{SC}|^2 \vert\psi_{\mathbf k}(z_0)\vert^2 G_{\sigma\tau;\sigma^{\prime}\tau^{\prime}}^{(SC)}({\mathbf k}, \omega; z_0), \label{SigmaSC}.
\end{equation}

Since the superconducting Hamiltonian $H_{\rm{SC}}$ is non-number conserving, the Green-function $G^{(\rm{SC})}$ contains both normal and anomalous
components that must be described in the Nambu-spinor notation. Therefore,
in addition to having spin-indices $\sigma\sigma^{\prime}$, the Green-function $G^{\rm{SC}}$ also has Nambu indices $\tau$ and $\tau^{\prime}$,
which take values $\pm 1$ similar to their spin counterparts $\sigma$ and
$\sigma^{\prime}$. Introducing the Pauli matrices $\sigma_{x,y,z}$ in the
spin components $\sigma,\,\sigma^{\prime}$ and the Nambu matrices $\tau_{x,y,z}$ in the
Nambu indices $\tau\,\tau^{\prime}$ allows us to write the Green-function
for the superconductor as
\begin{equation}
G^{(\rm {SC})-1}(\bm k;\omega)=(\epsilon_k-E_F)\tau_z+\Delta\tau_x-\omega.
\end{equation}
Substituting the Green-function into Eq.~\ref{SigmaSC} and assuming a slowly
varying density of states $\rho_{SC}(E_F)$ at the fermi-level $E_F$
in the superconductor allows us to simplify the self-energy as
\begin{equation}
\Sigma^{(\rm{SC})}({\mathbf k}, \omega) \approx  -\frac{-\omega+\Delta\tau_x}{\sqrt{\Delta^2-\omega^2}}, \label{SigmaSC1}
\end{equation}
for the energy levels of interest satisfying $\omega<\Delta$, where
$\lambda=|\tilde{t}_{SC}|^2 \vert\psi_{\mathbf k}(z_0)\vert^2\rho_{SC}(E_F) $.\cite{robustness,tudor,long}

The self-energy $\Sigma^{(\rm{SC})}$ induced in the Sm layer by integrating out the superconductor has an anomalous part (i.e proportional to $\tau_x$) and therefore induces a non-vanishing superconducting order parameter $\expect{\psi^\dagger_{Sm}\psi^\dagger_{Sm}}$ in the Sm
despite the fact that  the microscopic pairing potential
 $\Delta_{Sm}(\bm r)=0$ in the original Hamiltonian in Eqn.~\ref{Htot}.
 The microscopic pairing potential
 $\Delta_{Sm}(\bm r)=0$ vanishes in the semiconductor
since we assume no significant attractive pairing interaction in the semiconductor
 layer.
 Our assumption of a vanishing  $\Delta_{Sm}(\bm r)$  is also
consistent with the de Gennes boundary conditions
at the Sm-SC interface which requires $\Delta(\bm r)/N(\bm r) V(\bm r)$
 to be
continuous across the Sm-SC interface \cite{degennes}.
Here $\Delta(\bm r)$, $V(\bm r)$, $N(\bm r)$ stand for the microscopic
pairing potential, pairing interaction and density of states at the
fermi level on both sides of the interface. Since
$\Delta_{Sm}$ and $V_{Sm}$ are both zero on the Sm side of the interface,
the ratio can be finite which allows
 $\frac{\Delta_{Sm}}{V_{Sm}N_{Sm}}=\frac{\Delta_{SC}}{V_SC N_SC}$.

 Following the arguments in Ref.~\onlinecite{robustness}, to compute
the Green
function in the Sm layer, one can start by
integrating out the electronic degrees of freedom
 in the layers $F$ and $SC$ and replacing them with the
effective self-energies in Eq.~\ref{Sigmam} and Eq.~\ref{SigmaSC} respectively.
The Green function for $\omega\ll\Delta$ can then be obtained as
the inverse of an effective Hamiltonian for the low-energy excitations in
the Sm which is obtained by adding the respective self-energies from
the F and SC layers to Eq.~\ref{H0eff}  and can be written as,
\begin{equation}
H_{BdG}=-(\mu+\nabla^2)\tau_z-\imath\alpha(\bm\sigma\times\nabla)\cdot\hat{z}\tau_z+\Delta \tau_x+V_Z(x)\sigma_z,
\label{Hamiltonian1}
\end{equation}
where $\mu$ indicates the chemical potential in the Sm layer
 and $\bm\nabla=(\partial_x,\partial_y)$.
The effective pairing potential $\Delta$ and Zeeman potential $V_Z$
induced in the Sm layer is found to be proportional to $\Delta_{SC}$
and $V_F$ respectively, up to a renormalization that is dependent
on the tunneling strengths $\tilde{t}_{\rm{SC}}$ and $\tilde{t}_{\rm{F}}$
. \cite{long}

The Hamiltonian in Eq.~\ref{Hamiltonian1} is the effective Hamiltonian
of the F-Sm-SC sandwich structure which has recently been shown to
support a topologically superconducting phase in the appropriate parameter
regime.\cite{sau_prl'09} More specifically, it was shown that
when the parameters of the system are tuned so that
\begin{equation}
V_Z^2>\Delta^2+\mu^2
\end{equation}
the quasi-two-dimensional system would support zero-energy Majorana
fermions in vortices and chiral propagating Majorana modes around the
edges of the system. The chirality of the edge mode would be determined
by the sign of the Rashba spin-orbit coupling $\alpha$ and the Zeeman
splitting $V_Z$, the combination of which breaks the inversion and time-reversal symmetries of the system. (The explicit breaking of time-reversal
invariance due to the presence of the $V_Z$ term qualitatively 
distinguishes the topological superconductivity in the Sc-Sm-F sandwich 
structure \cite{sau_prl'09} from that in the SC-TI heterostructure 
\cite{fu_prl'09} which preserves time-reversal invariance.)

Finally, we would like to comment that the qualitative features of the results derived in this section for the
lowest semiconductor band carry over to the multi-band case \cite{long} and the topological
properties remain identical as well as long as an odd number of bands are occupied \cite{kitaev,roman-tudor1,roman-tudor2}.

\section{Chiral currents at magnetic domain walls}
Interferometry of Majorana fermions requires the ability to transport Majorana fermions along
two different paths and then recombining them.\cite{dassarma_prl'05,stern_prl'06,bonderson_prl'06,fu_prl'09,akhmerov_prl'09}
The semiconductor sandwich structure, in the appropriate parameter
regime, contains propagating chiral Majorana fermion modes.  However,
as mentioned in the introduction, the Majorana fermions in these
structures are neutral and therefore cannot be directly controlled by electrical voltages
or measured as electrical currents.
To avoid this difficulty, we will consider systems where a pair of isolated
chiral Majorana modes propagating in the same direction
are fused to form a chiral Dirac mode at a magnetic domain wall,
 which, as we will show,  carries a measurable electrical current.

 To model a domain-wall consisting of two
 isolated edges (marked $1$ and $2$ in Fig.~\ref{Fig1})
 in the semiconductor, we will take $V_z (x) > 0$ ($V_z(x) < 0$) for $x < -\frac{W}{2}$ ($x > \frac{W}{2}$) and $V_z(x)=0$ in between.
Opposite $V_z$ on the two sides of the edge can be realized by depositing
separate magnetic insulators with opposite magnetization.
 As shown in Ref.~[\onlinecite{sau_prl'09}], for $V_z=0$ the system is a (non-topological) regular $s$-wave superconductor. Therefore,
 in the geometry in Fig. 1 there are two isolated edges in the semiconductor each separating contiguous domains of non-Abelian
 $(V_Z^2>(\Delta^2+\mu^2))$ and regular $s$-wave superconducting phases.

Since the Hamiltonian is translationally symmetric along the $y$-direction,
we can factorize the energy eigenstates as $\psi_{k_y}(x,y)=e^{\imath k_y y}\phi_{k_y}(x)$ so that the solutions for $\phi_{k_y}(x)$ are determined by the $y$-independent Hamiltonian, obtained from Eq.~(\ref{Hamiltonian1})
by the substitution $\partial_y\rightarrow \imath k_y$.

The second-quantized current operator in the spin-orbit coupled semiconductor is given by \cite{macdonald},
\begin{align}
&\hat{\bm J}(\bm r)=\sum_{\beta}\imath\left[\bm\nabla\hat{c}_\beta^\dagger(\bm r)\hat{c}_\beta(\bm r)-\hat{c}_\beta^\dagger(\bm r)\bm\nabla\hat{c}_\beta(\bm r)\right]\nonumber\\
&+\alpha \sum_{\beta,\gamma}\hat{c}_\beta^\dagger(\bm r)\left(\hat{z}\times\bm \sigma\right)_{\beta\gamma}\hat{c}_\gamma(\bm r)
\label{eq:current}.
\end{align}
The current carried by the quasi-particle $\hat{\gamma}^\dagger$
 is then given by
\begin{align}
&\langle \hat{\bm J}\rangle_\gamma\equiv \langle\Psi|\hat{\gamma} \hat{\bm J}\hat{\gamma}^\dagger|\Psi\rangle-\langle\Psi|\hat{\bm J}|\Psi\rangle\label{eq:qpJ},
\end{align}
where $|\Psi\rangle$ is the ground state.
From
the above equations,
 the expression for the quasiparticle current along $y$ reduces to
\begin{equation}
\langle \hat{J}_y\rangle_\gamma=\int dx \phi_{k_y}^\dagger(x)(2k_y+\alpha\sigma_x)\phi_{k_y}(x).\label{eq:BdGcurrent}
\end{equation}
 In what follows, we will take the $y$-current operator
in the BdG notation to be $J_y=(2k_y+\alpha\sigma_x)$.

The magnetic domain wall composed of magnetic layers of opposite 
magnetization (Fig. 1) is the key idea underlying our proposed
Majorana interferometer since it enables an effective conversion 
of neutral Majorana modes into measurable charged current modes 
with current given by Eq.~\ref{eq:BdGcurrent}.

\section{Conductance in the wide domain-wall limit}

In the limit of $W\rightarrow\infty$, we can think of the interface
as being composed of two isolated edges at $\pm W/2$.
Zero-energy, $k_y=0$,  solution at each of these
edges is particle-hole symmetric, \emph{i.e.}, they satisfy
$\phi_j(x)=\sigma_y\tau_y\phi^*_j(x)$
where $j$ takes values $1$ or $2$, corresponding to
$4$-spinor wave functions at $-W/2$ $(\phi_1(x))$, and
at $W/2$ $(\phi_2(x))$, respectively.
The particle-hole transformation is defined as $\Xi=\sigma_y\tau_y K$
where $K$ is the complex conjugation operator.
Alternatively, the Majorana state $\phi_j$ can be written
in a manifestly particle-hole symmetric form
$\phi_j(x)=(u_j(x),-\imath\sigma_y u_j^*(x))^T$.
The solution $u_1(x)$ is real and inside the interface falls off as
\begin{equation}u_1(x)=u_1^*(x)=e^{-Re(z)x}[\rho e^{-\imath Im(z) x}+c.c]
\label{eq:asymp}
\end{equation}
 where $\rho$ is a constant 2-spinor and $z\approx (\frac{1}{\xi}+\imath k_F)$  with $\xi=\frac{v_F}{\Delta}$. Since $V_Z(x)=V_Z(-x)$, the Hamiltonian
commutes with $\sigma_x P$ where $P$ is the reflection operator with respect to the $x$ axis.  Hence, $\sigma_x P$ maps the Majorana mode $\phi_1(x)$ localized at $-W/2$ to the mode localized at $W/2$, \emph{i.e.}, $\sigma_x P\phi_1(x)=\sigma_x \phi_1(-x)=\lambda\phi_2(x)$ where $\lambda$ is a phase factor ($\lambda^*=\lambda^{-1}$).
After some straightforward algebra, the requirement that both $\phi_1(x)$ and $\phi_2(x)$ are particle-hole symmetric
requires $\lambda^2=-1$ or $\lambda=\imath$.
Using $k.p$ perturbation theory, the solutions $\phi_{j,k_y}(x)$ at
 $k_y\neq 0$ can be
 approximated to lowest order in $k_y$
by $\phi_{j,k_y}(x)\approx\phi_j(x)$ and have energy $\epsilon_k=v_g k_y$ where
\begin{equation}
v_g=\alpha\langle\phi_j|\sigma_x\tau_z|\phi_j\rangle=2\int dx u_j^\dagger(x)\sigma_x u_j(x)
\end{equation}
is the group velocity of the modes.

The BdG current operator (Eq.~\ref{eq:BdGcurrent}) for the Majorana
states around $k_y=0$ is given by $J_y=\alpha\sigma_x$.
For $W\rightarrow \infty$, we get the current
carried by the Majorana states $\phi_j$ to be
\begin{align*}
&\langle\phi_j|J_y|\phi_j\rangle=\alpha\int dx u_j^\dagger \sigma_x u_j +
u_j^T \sigma_y\sigma_x\sigma_y u_j^*\nonumber\\
&=\alpha\int dx u_j^\dagger \sigma_x u_j -u_j^\dagger \sigma_x u_j=0,
\end{align*}
as expected.

For finite $W$, the wave functions $\phi_j$ are no longer orthogonal and have finite overlaps. Following Ref.~[\onlinecite{cheng}], the lowest energy eigenstates for a finite-width interface can be
 considered to be superpositions of $\psi_{j,k_y}(x,y)=e^{\imath k_y y}\phi_j(x)$. Since $\sigma_x P$
commutes with the Hamiltonian,
these eigenstates should be $\psi_{\pm,k_y}(x,y)=e^{\imath k_y y}\zeta_{\pm}(x)$
such that $\sigma_x P\zeta_{\pm}(x)=\pm\zeta_{\pm}(x)$.
Using $\phi_2(x)=\lambda^{-1}\sigma_x \phi_1(-x)$
in the above it follows that
\begin{equation}
\zeta_{\pm}(x)=\frac{1}{\sqrt{2}}\left[\phi_1(x)\pm\imath \phi_2(x)\right].
\end{equation}
 The expectation value of an operator $A$ with respect to these energy
eigenstates, to lowest order in the overlap, is given by
\begin{eqnarray}
\langle\zeta_{\pm}|A|\zeta_{\pm}\rangle &=&\frac{\sum_j\langle\phi_{j}|A|\phi_{j}\rangle\pm 2 Im(\langle\phi_{1}|A|\phi_{2}\rangle)}{2 \pm 2 Im(\langle\phi_{1}|\phi_{2}\rangle}\nonumber\\&\approx& \frac{1}{2}\sum_j\langle\phi_{j}|A|\phi_{j}\rangle \pm Im(\langle\phi_{1}|A|\phi_{2}\rangle)\label{eq:expectation}.
\end{eqnarray}

Using the above formula we can compute the excitation spectrum of a finite-width interface as
\begin{equation}
\epsilon_{\pm}(k_y)=\langle\zeta_{\pm}|H_{BdG}(k_y)|\zeta_{\pm}\rangle=v_g k_y \pm \alpha M
\end{equation}
where $M$ can be written as \cite{bardeen},
\begin{equation}
M=\partial_x\left[\phi_1^\dagger(x)\sigma_x\tau_z\phi_1(x)\right]_{x=0}=2\partial_x \left[u_1^\dagger(x)\sigma_x u_1(x)\right]_{x=0}
\end{equation}
Thus, for finite $W$, tunneling splits the pair of chiral Majorana modes into a pair of
 modes with dispersion $\epsilon_{\pm}(k_y)=v_g(k_y\pm k_1)$
where $k_1=\alpha M/v_g$.

The states $\zeta_{\pm}$, unlike their constituent states $\phi_j$,
can carry a charge current.
The  expectation value of the current operator in these
states is given by the
 cross-term
\begin{align}
&\alpha\langle\phi_1|\sigma_x|\phi_2\rangle=\alpha\int dx u_1^\dagger \sigma_x u_2 +
u_1^T \sigma_y\sigma_x\sigma_y u_2^*\nonumber\\
&=2\imath \alpha\int dx Re(u_1^\dagger(x) u_1(-x)),
\end{align}
which is now non-zero as advertised in the introduction. Let us now estimate
the conductance of the interface using the states $\zeta_\pm$
\cite{btk'82}.
The branches $\psi_{\pm,k_y}(x,y)$ are related by
 particle-hole conjugation. Therefore it suffices to consider the
 occupancy of only the electron-like branch defined by  $\langle\tau_z\rangle>0$.
Only the sign of the current
 $\langle \hat{J}_y\rangle$ depends on which of the two particle-hole
 conjugate branches is electron-like.
 Therefore, we assume the $+$ branch to be electron-like and multiply the
result by  the sign of $\langle\tau_z\rangle_+=2\int dx u_1^T(-x)\sigma_x u_1(x)$ at the end of the calculation.
 For a finite bias voltage $V$, which is applied to the interface
 by coupling it to a reservoir at a chemical potential $(\mu_S+e V)$
 via a tunnel barrier, the occupation function of the  mode is given by
$f(\epsilon_{+}(k_y)-e V)$ where $f(\epsilon)$ is the Fermi function \cite{btk'82}.
The conductance of the interface for $V\rightarrow 0$ can be computed using a
Landauer-like formula
\begin{align}
&G(0)=\frac{d}{d V}\int d k_y f(\epsilon_{+}(k_y)-e V) \langle \hat{J}_y\rangle_{+,k_y}|_{V=0}=\frac{\langle \hat{J}_y\rangle_{+,-k_1}}{v_g}G_0
\end{align}
where $G_0$ is the quantum of conductance.
Substituting the explicit form for the expectation value with respect to
 the
energy eigen-modes (Eq. \ref{eq:expectation}), $G(0)$ can be written as
\begin{equation}
G(0)= \frac{2 G_0}{v_g}\sgn{(\langle\tau_z\rangle_+)}\left[-k_1+\int dx Re(u_1^\dagger(x) u_1(-x))\right].
\end{equation}
Using the asymptotic form of the chiral edge state wave-functions (Eq.~(\ref{eq:asymp})), the above
expression, in the large $W$ limit, can be shown to be of the form
\begin{equation}
G(0)\sim \sgn{(\langle\tau_z\rangle_+)} e^{-Re(z)W}\cos{(2 Im(z)W+\delta)}G_0
\end{equation}
where $\delta$ is a phase-shift.
As claimed in the introduction, this is finite for finite interface width,
 but vanishes exponentially with $W$ with a decay length $Re(z)^{-1}=\xi$.

\section{Numerical solution for narrow domain-walls}
For experimental purposes, it is
interesting to consider the limit where the width of the interface $W$ is much smaller than the coherence length $(W\ll\xi)$ of the
 superconductor. For definiteness we approximate this limit as $W=0$,
 in which case there is a single interface at $x=0$.

As in the wide-interface case, the zero-bias
conductance can be written in terms of the $E=0$ energy eigenstate $\psi_{+,-k_1}(x,y)=e^{-\imath k_1 y}\zeta_{+}(x)$, which is determined by its value
on one side of the interface, say $x>0$, where it can be expanded as $\zeta_+(x)=\sum_{n:Re(z_n)<0}c_n e^{z_n x}\phi_n.$
Here $\phi_n$ and $z_n$ are complex eigenvectors and eigenvalues of the Hamiltonian $H_{BdG}(k_y,z)$ in Eq.~[\ref{Hamiltonian1}] (with the substitutions $\partial_x\rightarrow z$ and $\partial_y\rightarrow \imath k_y$),\emph{ i.e.}, they satisfy $H_{BdG}(k_y,z_n)\phi_n=0.$
 The boundary conditions at $x=0$ are then given by
$\sigma_x \zeta_{+}(0)=\zeta_{+}(0)$ and $\sigma_x \zeta_{+}'(0)=-\zeta_{+}'(0)$.
We vary $k_y$ on the real axis to find the value $k_y=-k_1$
where the boundary
conditions on the wave function are satisfied. The
expectation value of an observable $A$ is calculated as
\begin{equation}
\langle A\rangle=\int dx \zeta_{+}^\dagger(x)A\zeta_{+}(x)=\sum_{n,m} c_n^*c_m\frac{\phi_n^\dagger(A+\sigma_x A\sigma_x)\phi_m}{z_m+z_n^*}.
\end{equation}
 We numerically calculate $c_n$
for a  representative set of parameter values
appropriate for the non-Abelian superconducting phase: $\Delta=0.5,\,\, \mu=0.2,\, \alpha=1.0$ in units where all energies are scaled by  $V_Z=1$ meV
and lengths are scaled by $\hbar/\sqrt{2 m^* V_Z}$ \cite{sau_prl'09}. In this calculation, the conductance in the limit $V \rightarrow 0$ is found to be $G(0)=0.88 G_0$.
To avoid decoherence processes such as scattering of electrons into holes,
we find that $V,T<k_1 v_g\sim 0.1$ meV where $T$ is the temperature.

\section{Domain-wall based majorana fermion interferometers}
Since the interface between two domains of the semiconductor heterostructure with opposite directions of Zeeman splitting
carries a charge current, constructing electrically controlled Majorana interferometry experiments \cite{fu_prl'09,akhmerov_prl'09} is straightforward.
In Fig. 2a, the interfaces $a$ and $d$ carry charge current.
In contrast to interfaces $a$ and $d$, the edges $b$ and $c$ separate a non-Abelian superconducting phase from the vacuum in the central hole.
Hence, they carry charge-neutral Majorana modes which we denote by $\gamma_{0b}, \gamma_{0c}$.
At the tri-junction between the edges $a, b,$ and $c$, a chiral charged quasiparticle or quasihole in interface $a$ breaks into a pair of Majorana fermions in $b$ and $c$: $c_a^{\dagger}\rightarrow \gamma_{0b}+i\gamma_{0c}, c_a\rightarrow \gamma_{0b}-i\gamma_{0c}$.There are an even (including zero) or odd number $n_V$ of flux quanta threaded into the central hole in the interferometer.
 Consequently,
for an odd (even) number of vortices in the central hole, only one (none) of $\gamma_{0b}$ and $\gamma_{0c}$
 encounters a change of sign while traversing the edges $b$ and $c$.
It follows that, for an odd number of vortices in the central hole, $c_a^{\dagger}\rightarrow c_d$ and $c_a\rightarrow c_d^{\dagger}$ (a quasiparticle incident from $a$ turns into a quasihole at $d$ or vice versa), and charge $2e$ is transferred between the interface $a$ and the superconductor. Therefore, if the number of vortices in the hole is odd (even), there is a non-zero (zero) current between the interface $a$ and the superconductor, which is a definitive signature of non-Abelian statistics of the Majorana quasiparticles.

We now consider an interferometer geometry capable of measuring the fermion number in a
topological qubit made of two vortices (Fig 2b). Such a measurement is important for
 TQC in the semiconductor.\cite{bravyi1,bravyi2,bonderson_ann_phys}
\begin{figure}
\centering
\includegraphics[scale=0.3,angle=0]{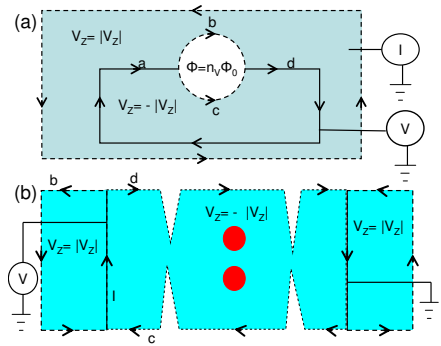}
\caption{(a): Interferometer geometry to test non-Abelian statistics of Majorana fermions.
There is a non-zero current $I$ flowing in the superconductor when the number of vortices
$n_V$ in the central hole is odd. $\phi_0$ indicates the superconducting flux quantum
$\frac{hc}{2e}$; (b): Interferometric
measurement of the fermion number in a pair of vortices shown as (red) circles forming a qubit at the center.
Charged fermion quasiparticles carrying current $I$ break into Majorana modes along edges $b$ and $d$,
which can recombine after quantum tunneling at the two edge constrictions on the two sides of the qubit.}
\end{figure}
Charged fermion quasiparticles carrying current $I$ at the left interface break up
into two Majorana fermions at edges $b$ and $d$. The Majorana fermion at $d$ can quantum mechanically tunnel to edge $c$
via two edge constrictions on the two sides of the central qubit. We assume that the constrictions
have small enough capacitance so that the Majorana fermion tunneling amplitudes $t_1, t_2$ via \emph{vortex tunneling} (quantum phase slips)
 are appreciable in the background of superconducting quasiparticle tunneling. Writing the net such tunneling amplitude as $t_{dc}$ and the number of fermions trapped in the central qubit as $n_f$,
 we get\cite{fradkin},
 \begin{equation}
 t_{dc}\propto t_1^2 + t_2^2 + (-1)^{n_f}2t_1t_2.
 \label{interference}
 \end{equation}
 Here, the factor of $(-1)^{n_f}$ arises from the phase picked up by a vortex on one full circle around $n_f$ fermions.
It follows that for $t_1\sim t_2$ the contribution to the current
from quantum phase slips at the two constrictions vanishes for an odd number of fermions at the center.
Therefore, with change in $n_f$ at the central qubit, the magnitude of the current $I$ shows
well defined oscillations.
We note that quantum vortex tunneling, i.e.  quantum phase slip,
is an essential ingredient of our interferometer. 
\section{Conclusion}

 The electrically controlled edge-state interferometry on the semiconductor heterostructure we propose in this paper are
 exactly analogous to those proposed previously for the non-Abelian states in the
 $\nu=5/2$ fractional quantum Hall system and on the surface of a 3D strong topological insulator. \cite{dassarma_prl'05,stern_prl'06,bonderson_prl'06,fu_prl'09,akhmerov_prl'09}
 The Majorana fermion edge excitations on the fractional quantum Hall systems are charged \cite{dassarma_prl'05,stern_prl'06,bonderson_prl'06}
and charged
 edge modes have also been shown to be possible on the surface of a 3D topological insulator. \cite{fu_prl'09,akhmerov_prl'09} Even though the
  charge of the edge excitations make electrically controlled interferometry fairly straightforward in these systems, there
  have been lingering questions if such experiments are possible on the semiconductor structure because of the absence
  of charge of the chiral modes on any isolated edge in this system. In this paper we have shown that two isolated edges
  carrying individually charge-less Majorana edge modes can be \emph{superposed} to create an edge which does carry charge. The charge
  of this new edge mode can be used to electrically control edge currents which in turn can be used in the interferometry experiments.
Quantum phase slips, which are essential for our scheme to work have 
been experimentally observed, \cite{vlad} and therefore, our scheme should 
work as a matter of principle.

There is a natural sequence of experiments in the study of Majorana modes 
in the semiconductor sandwich structures. The first experiments, which 
are already underway, \cite{expts} would fabricate the 
appropriate SC/Sm heterostructures searching for the proximity-induced 
superconductivity in the appropriate parameter space of chemical 
potential and Zeeman splitting. Once superconductivity is identified 
in the semiconductor, suitable tunneling spectroscopic measurements 
would have to establish a robust zero-energy anomaly in the semiconductor 
consistent with the existence of the zero-energy Majorana mode. 
One great advantage of the sandwich structure is the generic existence of 
the Majorana mode by construction in the topological superconducting 
phase in contrast to the 5/2 FQHE state where the non-Abelian ground
state is a conjecture. The observation of the zero-bias anomaly showing 
the existence of a zero-energy state in the BdG spectrum would satisfy the necessary condition for the existence of the Majorana mode, but not the 
sufficient condition. The proposed fractional Josephon effect 
experiment \cite{kitaev,roman} would provide the sufficient 
condition, but direct interferometry, as proposed in the current 
work, is essential in establishing the non-Abelian nature of the 
Majorana particles. What we propose here is not by any means an easy 
experiment, but it is probably not much harder than the on-going 
interferometric measurements \cite{willett} in the 5/2 FQHE state.
We emphasize that only interferometry can directly establish the 
existence of non-Abelian braiding statistics, and as such our 
proposed experiment is an important necessary step in the 
beautiful and exotic physics of emergent solid state non-Abelian 
modes. 

  Our proposed interferometry experiments give clear cut signatures of the non-Abelian statistics of the Majorana modes localized in the
  order parameter defects on the semiconductor heterostructure. In addition to the zero-bias conductance peak experiments \cite{long}
  and the experiments measuring the fractional Josephson effect \cite{roman,oreg} due to Majorana fermions, these experiments constitute
  another class of measurements directly testing the existence of non-Abelian quasiparticles in the semiconductor heterostructure.
  Moreover, the interferometry experiments, like their counterparts in the $\nu=5/2$ state and on the surface of a topological insulator,
  constitute the necessary first step in entanglement generation and implementation of TQC using the 2D semiconductor heterostructure.


We thank Roman Lutchyn for discussions.
This work is supported by DARPA-QuEST, JQI-NSF-PFC, and LPS-NSA. ST acknowledges DOE/EPSCoR Grant \# DE-FG02-04ER-46139 and Clemson University start up funds for support.





\end{document}